\documentclass[aps,prb,twocolumn,showpacs,superscriptaddress,letterpaper]{revtex4}
\usepackage{graphicx,epsfig,amsmath,amssymb,amsfonts,mathrsfs,psfrag,color}
\usepackage{mathrsfs,mathtools,relsize,bm}
\pdfoutput=1

\newcommand{\al}{\alpha}
\newcommand{\be}{\beta}
\newcommand{\g}{\gamma}
\newcommand{\de}{\delta}
\newcommand{\e}{\epsilon}

\newcommand{\ka}{\kappa}
\newcommand{\la}{\lambda}

\newcommand{\p}{\pi}

\newcommand{\s}{\sigma}

\newcommand{\f}{\phi}
\newcommand{\x}{\chi}

\newcommand{\W}{\Omega}
\newcommand{\De}{\Delta}

\newcommand{\round}[1]{\left( #1 \right)}
\renewcommand{\square}[1]{\left[ #1 \right]}
\newcommand{\curly}[1]{\left\{#1\right\}}
\newcommand{\abs}[1]{\left| #1 \right|}

\newcommand{\mat}[4]{\left(\begin{array}{cc}#1&#2\\#3&#4\end{array}\right)}

\newcommand{\cvecf}[4]{\round{\begin{array}{c}#1\\#2\\#3\\#4\end{array}}}

\newcommand{\ang}[1]{\left\langle #1 \right\rangle}

\newcommand{\bra}[1]{\left\langle #1\right|}
\newcommand{\ket}[1]{\left| #1\right\rangle}

\newcommand{\beq}{\begin{equation}}
\newcommand{\eeq}{\end{equation}}
\newcommand{\Beq}{\begin{align}}
\newcommand{\Eeq}{\end{align}}
\newcommand{\bml}{\begin{multline}}

\newcommand{\bsp}{\begin{split}}
\newcommand{\esp}{\end{split}}
\newcommand{\down}{\downarrow}
\newcommand{\up}{\uparrow}

\newcommand{\nn}{\nonumber}

\newcommand{\bk}{{\bf k}}
\newcommand{\bq}{{\bf q}}

\newcommand{\hH}{\hat H}
\newcommand{\bH}{{\bf H}}

\newcommand{\bQ}{{\bf Q}}
\newcommand{\bb}{{\bf b}}
\newcommand{\bbb}{\bar{\bf b}}

\newcommand{\cG}{\check{\mathscr{G}}}

\newcommand{\bB}{{\bf B}}

\newcommand{\cH}{\check H}
\newcommand{\cV}{\check V}

\newcommand{\hI}{\hat{\mathbb{I}}}
\newcommand{\cI}{\check{\mathbb{I}}}
\newcommand{\hpmb}{\hat{\bm\s}}
\newcommand{\hbbpm}{\hat{\bar{\bm\s}}}
\newcommand{\hpm}{\hat{\s}}
\newcommand{\bka}{\bm{\ka}}

\DeclareMathOperator{\tr}{Tr}

\DeclareMathOperator{\diag}{Diag}
\newcommand{\ev}[1]{{\left\langle{#1}\right\rangle}}

\newcommand{\rr}{\mathbf{r}}

\newcommand{\KK}{\mathbf{K}}
\newcommand{\bn}{\mathbf{n}}
\newcommand{\dm}{\hat\varrho}

\begin{document} 

\title{Low-density molecular gas of tightly-bound Rashba-Dresselhaus fermions}
\author{So Takei}
\affiliation{Condensed Matter Theory Center, Department of Physics, The University of Maryland College Park, MD, 20742, USA}
\author{Chien-Hung Lin}
\affiliation{Condensed Matter Theory Center, Department of Physics, The University of Maryland College Park, MD, 20742, USA}
\author{Brandon M. Anderson}
\affiliation{National Institute for Standards and Technology, Gaithersburg, MD, 20899, USA}
\author{Victor Galitski}
\affiliation{Condensed Matter Theory Center, Department of Physics, The University of Maryland College Park, MD, 20742, USA}
\affiliation{Joint Quantum Institute, Department of Physics, The University of Maryland College Park, MD, 20742, USA}
\date{\today}

\begin{abstract}
We study interacting Rashba-Dresselhaus fermions in two spatial dimensions. First, we present a new exact solution to the two-particle pairing problem of spin-orbit-coupled 
fermions for arbitrary Rashba and Dresselhaus spin-orbit interactions. An exact molecular wave function and the Green function
are explicitly derived along with the binding energy and the spectrum of the molecular state. 
In the second part, we consider a thermal Boltzmann gas 
of fermionic molecules and compute the time-of-flight velocity and spin distributions for a single fermion in the gas. We show that the 
pairing signatures can be observed already in the first-moment expectation 
values, such as time-of-flight density and spin profiles.
\end{abstract}

\maketitle

\section{Introduction}
The concept of bound states of fermion pairs arises ubiquitously in physics. The interior of a neutron star is believed to be a superfluid of 
bound neutron pairs held together by an attractive component of the inter-nucleon interactions.\cite{sauls} In particle physics, 
mesons are a bound state of two fermions, a quark and an anti-quark, with the attraction provided by the strong interaction.
Arguably the most well-known two-fermion bound state in condensed matter physics is the Cooper pair\cite{cooper}, bound together in some cases 
through phonon exchange\cite{BCS}, which is the key component in the understanding of superconductivity.

A bound state of a fermion pair also plays an important role in the field of ultracold gases of neutral Fermi atoms, where  
atoms in two different hyperfine states can be used to realize pairing. With the advent of the Feshbach resonance\cite{feshbach1,
feshbach2,feshbach3,feshbach4,feshbach5} experimentalists 
are now able to adjust both the magnitude and sign of the two-body scattering length by simply tuning an external magnetic field\cite{feshbach6,
feshbach7,feshbach8,feshbach9,feshbach10}. This has opened 
doors to a new field which explores the crossover between a BCS superfluid of weakly bound Cooper pairs and a Bose-Einstein condensate 
(BEC) of tightly bound fermionic molecules.\cite{becbcs1,becbcs2,becbcs3,becbcs4} 
The BEC of such molecules have been observed in time-of-flight (TOF) 
experiments which show a characteristic bimodal distribution similar to the one observed in a BEC of bosonic atoms.\cite{bimodal1,bimodal2,
bimodal3,bimodal4,bimodal5}

In addition to the unprecedented control over the strength and nature of inter-atomic interactions experimentalists are also capable of
fine-tuning with relative ease the structures and parameters of these atomic systems using quantum optical techniques. These developments
open the possibility of emulating various solid state systems using ultra-cold atomic systems, and using them to gain insights into the 
outstanding problems in strongly correlated condensed matter physics.\cite{ucfermi,rev1,rev2} In this regard, the theoretical proposals\cite{gauge1,
gauge2,gauge3,gauge4,gauge5,gauge6} for generating 
synthetic Abelian and non-Abelian gauge fields for a gas of neutral fermionic atoms, and the recent experimental realization of the phenomenon by
the NIST group\cite{nist1,nist2,nist3}, brings great excitement to the community. While a spatially inhomogeneous Abelian gauge field can mimic 
the role of an
external magnetic field, a uniform non-Abelian gauge field is capable of introducing spin-orbit (SO) interactions akin
to Rashba and Dresselhaus couplings found in semiconductor electronic systems.\cite{gauge5} These achievements may serve as a key ingredient when
realizing some of the most intriguing phases of matter in condensed matter physics, including fractional quantum Hall liquids\cite{fqh1,fqh2} and 
topological insulators\cite{ti1,ti2}.

A fundamental feature of low-density electronic systems with Rashba SO interaction is the enhanced
tendency for bound pair formation due to a SO-induced increase in the density of states of low-energy electrons.\cite{socenhance} 
An inspection of the
single-particle energy spectrum for a Rashba particle reveals a ring of minima in energy in contrast to a single minimum at the 
momentum-space origin for a particle without SO coupling. Indeed, it has been noted, for instance, that SO interaction can be remarkably 
beneficial for superconducting pairing.\cite{marsiglio}
The experimental realization of SO interactions in ultra-cold Fermi gases has motivated many recent theoretical works.
\cite{bcsbecsocrev} Some of these include the BCS-BEC crossover physics\cite{bcsbecsoc1,bcsbecsoc2,bcsbecsoc3,bcsbecsoc4,bcsbecsoc5} 
and the effects of population imbalance and superfluidity\cite{sf1,sf2,sf3,sf4,sf5,sf6} in the presence of SO interactions, as well as
properties of the BEC ground state in the strong gauge-field limit\cite{rashbons}. 

In this work, we study tightly-bound molecules of Rashba-Dresselhaus fermions in two spatial dimensions,
and investigate the properties of a low-density thermal gas of such molecules. We first develop in detail the exact solution to the
problem of two Rashba-Dresselhaus fermions interacting via an attractive short-ranged $s$-wave interaction. The wave function for the bound molecular 
state is then obtained by solving for the bound state energy and the molecular energy spectrum. We then focus on a low-density Boltzmann gas 
of these bound molecules and compute the \textit{single-fermion} density matrix 
for one of the two fermions that form a molecule in the gas. With this density matrix we evaluate the fermion velocity 
and spin density distributions in momentum space that can be inferred from a TOF experiment. We compare and contrast
these distributions with those corresponding to a non-interacting Boltzmann gas of SO-coupled fermions. 
The main finding that stems from this consideration is that the pairing of a fermion with its molecular partner has an 
experimentally observable imprint on these distributions, and that these imprints are absent when fermions are not interacting. 
We show that these signatures of pairing appear already, for instance, in the first-moment density expectation value
$\ang{n(\bk)}$. Higher-order correlations as a means to probe many-body states has previously been proposed in the context of 
ultra-cold atoms.\cite{demler}

The paper is organized as follows. The exact solution to the above-mentioned interacting two-particle problem is presented in Sec. 
\ref{exactsol}. The properties of the bound molecular state are explored in Sec. \ref{molstate}: analytic expressions for the binding
energy and the molecular spectrum in different physical limits are shown in Secs. \ref{be} and \ref{spectrum}, respectively. Effects due to
a Zeeman term is briefly discussed in Sec. \ref{zeeman}. In Sec. \ref{probe}, we consider a thermal gas of non-interacting molecules
and focus on the single-fermion density matrix for such a system (Sec. \ref{quantitative}). There, we compute the TOF velocity (Sec. \ref{sec:veldist})
and spin (\ref{sec:spindensity}) distributions for a single fermion in the system. In Sec. \ref{experiments} we discuss how these distributions
can be measured experimentally.

\section{The Exact two-particle wave function}
\label{exactsol}
Let us begin with a system of two ``spin"-1/2 fermions with two-dimensional Rashba and Dresselhaus SO interactions moving in two
spatial dimensions. The single-particle Hamiltonian is given by
\beq
\label{HRD}
\hH_{RD}(\bk)=\e_\bk\hI+\al\round{\hpm_xk_y-\hpm_yk_x}+\be\round{\hpm_xk_x-\hpm_yk_y},
\eeq
where $\e_\bk=k^2/2m$ with $m$ being the mass of the particle,
$\hpmb=(\hpm_x,\hpm_y)$ are the usual Pauli matrices and $\bk=(k_x,k_y)$ is the particle momentum. We will be
using units where $\hbar=1$ throughout.
The term proportional to $\al$ ($\be$) is known as the Rashba (Dresselhaus) SO interaction. The hats denote $2\times2$ 
matrices acting on the ``spin"-space of the particle. Here, the ``spin" degree of freedom may be a \textit{synthetic} degree of freedom. 
Hamiltonian (\ref{HRD}), for instance, can be realized in ultra-cold atomic
systems where atoms with multiple internal levels move in the presence of a spatially modulated laser field.\cite{gauge5,nlevel}



For the two-particle system the kinetic energy contribution to the Hamiltonian is a $4\times4$ matrix given by
\beq
\label{H0}
\check H_0=\hH_{RD}(\bk_1)\otimes\hI+\hI\otimes\hH_{RD}(\bk_2),
\eeq
where $\bk_i$ with $i\in\{1,2\}$ is the momentum of the $i$-th particle and the check indicates an operator acting on the two-particle 
Hilbert space. For some interaction Hamiltonian $\cH_{\rm int}$ 
the total Hamiltonian of the system is then given by $\check H=\check H_0+\check H_{\rm int}$.

The energy eigenstates of our two-particle system are four-component spinors and are labeled by two indices, $\bQ$ and $n$, 
where $\bQ=\bk_1+\bk_2$ is the centre-of-mass momentum and $n$ labels both the various bound and scattering states. 
Any energy eigenstate $\ket{\psi_n(\bQ)}$ can be written as a superposition of momentum eigenstates as
\beq
\label{EE}
\ket{\psi_n(\bQ)}=\int\frac{d^2\bk}{(2\p)^2}\x_n(\bQ,\bk)\ket{\bQ,\bk},
\eeq
where $\bk=(\bk_1-\bk_2)/2$ is the relative momentum.
Note here that $\x_n(\bQ,\bk)$ are four-component spinors owing to the different possible spin
states of the two-particle system. We will be using the triplet-single basis, $\ket{S,s_z}=\{\ket{1,+1},\ket{1,-1},\ket{10},\ket{00}\}$,
to express the spinor (see (\ref{an2})), unless otherwise stated. The stationary Schr\"{o}dinger equation is then
\beq
\label{SE}
[\cH-\cI E_n(\bQ)]\ket{\psi_n(\bQ)}=0,
\eeq
where $E_n(\bQ)$ is the energy of the $(\bQ,n)$ state. Inserting (\ref{EE}) into (\ref{SE}), we obtain
\begin{multline}
\label{SE2}
\square{\cH_0(\bQ,\bk)-\cI E_n(\bQ)}\x_n(\bQ,\bk)\\
+\int\frac{d^2\bk'}{(2\p)^2}\bra{\bk}\cH_{\rm int}\ket{\bk'}\x_n(\bQ,\bk')=0.
\end{multline}
We note that a general isotropic interaction in two dimensions can be written as
\beq
\label{Hint}
\bra{\bk}\cH_{\rm int}\ket{\bk'}=\sum_{l=-\infty}^\infty\cV_l(k,k')e^{il(\f_\bk-\f_{\bk'})},
\eeq
where $\f_\bk$ is the angle between $\bk$ and the $x$-axis. We will hereafter assume that the interaction is short-ranged.
At low energies and long wavelengths, scattering amplitude is dominated by the contribution of the $s$-wave 
scattering as long as the relative momentum $\bk$ satisfies $kR_e\ll 1$, where $R_e$ is the characteristic
radius of the interaction. Due to the anti-symmetry of the two-particle wave function we must project out the singlet 
component of the wave function for the $s$-wave component (i.e. $l=0$). Dropping higher order harmonic contributions, we 
may then write the potential as $\bra{\bk}\cH_{\rm int}\ket{\bk'}\rightarrow V_0\check{\mathscr{P}}^{(s)}$, where $\check{\mathscr{P}}^{(s)}
=\ket{00}\bra{00}$ is the singlet projector. We are then led to rewriting (\ref{SE2}) as
\begin{multline}
\square{\cH_0(\bQ,\bk)-\cI E_n(\bQ)}\x_n(\bQ,\bk)\\
+V_0\int\frac{d^2\bk'}{(2\p)^2}\check{\mathscr{P}}^{(s)}\x_n(\bQ,\bk')=0.
\end{multline}

Let us now define the inverse Green function as $\cG_n^{-1}(\bQ,\bk)=\cH_0(\bQ,\bk)-\cI E_n(\bQ)$ and the function
$c^n_\bQ$ via
\beq
\label{cnQ}
c^n_\bQ\ket{00}=\int\frac{d^2\bk'}{(2\p)^2}\check{\mathscr{P}}^{(s)}\x_n(\bQ,\bk').
\eeq
Then we obtain the expression for the spinor, 
\beq
\label{an}
\x_n(\bQ,\bk)=-c^n_\bQ V_0\cG_n(\bQ,\bk)\ket{00}.
\eeq
The task now is to invert the inverse two-particle Green function, 
$\cG_n^{-1}(\bQ,\bk)$. Hereafter, we will refrain from writing the index $n$ explicitly since
it enters only to label the eigenenergies. We now introduce the single-particle inverse Green functions for particle $i$, namely, 
$\hat g^{-1}(\bk_i,\bQ)=\hH_{RD}(\bk_i)-\hat{\mathbb{I}}(E_n(\bQ)/2)=\hat{\mathbb{I}}y(\bk_i,\bQ)+\al\hat{\bm\s}\cdot(\bb_i\times\hat z)$, 
where $y(\bk_i,\bQ)=\e_{\bk_i}-E_n(\bQ)/2$ and
\beq
\bb_i=\mat{1}{\g}{\g}{1}\bk_i.
\eeq
Here, $\g=\be/\al$ measures the relative strength between the Rashba and Dresselhaus couplings.
The two-particle inverse Green function is then
\beq
\cG^{-1}(\bQ,\bk)=\square{\hat g^{-1}(\bk_1,\bQ)\otimes\hat{\mathbb{I}}+\hat{\mathbb{I}}\otimes\hat g^{-1}(\bk_2,\bQ)}.
\eeq
$\hat g^{-1}(\bk_i,\bQ)$ can be diagonalized using the unitary matrix
\beq
\label{ui}
\hat u_i=\exp\curly{-i\frac{\p}{4}\round{\bn_i\cdot\hat{\bm\s}}},
\eeq
where $\bn_i=\bb_i/b_i$. We then obtain $\hat u^\dag_i\hat g^{-1}(\bk_i,\bQ)\hat u_i=\hI y_n(\bk_i,\bQ)+\al b_i\hpm_z$. 
Therefore, $\cG^{-1}(\bQ,\bk)$ can be inverted using the composite unitary transformation $\check U=\hat u_1\otimes
\hat u_2$ and we obtain
\beq
\label{GFmat}
\cG(\bQ,\bk)=\check U\check D(\bQ,\bk)\check U^\dag
\eeq
with 
\begin{multline}
\label{diagmat}
\check D(\bQ,\bk)=\diag\{[D_1(\bQ,\bk)]^{-1},[D_2(\bQ,\bk)]^{-1},\\
[D_3(\bQ,\bk)]^{-1},[D_4(\bQ,\bk)]^{-1}\},
\end{multline}
and
\begin{align}
\label{D1}
D_1(\bQ,\bk)&=s+\al(b_1+b_2)\\
D_2(\bQ,\bk)&=s+\al(b_1-b_2)\\
D_3(\bQ,\bk)&=s-\al(b_1-b_2)\\
\label{D4}
D_4(\bQ,\bk)&=s-\al(b_1+b_2).
\end{align} 
Here, we have defined
\beq
\label{mmdefs}
s:=s(\bQ,\bk)=\frac{k^2}{m}+\frac{Q^2}{4m}-E_n(\bQ),
\eeq
An explicit expression for the Green function
matrix (\ref{GFmat}) is presented in Appendix \ref{GFexp}. Inserting (\ref{GFmat}) 
into (\ref{an}) the spinor $\x(\bQ,\bk)$ can be expressed in the triplet-singlet basis as
\beq
\label{an2}
\x(\bQ,\bk)=-\frac{c_\bQ V_0}{d(\bQ,\bk)}\cvecf{\mathscr{A}_{1}^t(\bQ,\bk)}{-\mathscr{A}_{1}^{t*}(\bQ,\bk)}
{\mathscr{A}_{0}^t(\bQ,\bk)}{\mathscr{A}^s(\bQ,\bk)}.
\eeq
The denominator $d(\bQ,\bk)$ is given by 
\beq
\label{denom}
d(\bQ,\bk)=s^4-4\al^2s^2\round{b^2+\frac{B^2}{4}}+4\al^4(\bB\cdot\bb)^2,
\eeq
where $\bB=\bb_1+\bb_2$, $\bb=(\bb_1-\bb_2)/2$.
The coefficients for the triplet and singlet components are explicitly given by
\begin{align}
\mathscr{A}_{1}^t(\bQ,\bk)&=i\sqrt{2}\square{s\al be^{-i\f_b}-\al^3(\bB\cdot\bb)Be^{-i\f_B}}\\
\mathscr{A}_{0}^t(\bQ,\bk)&=2is\al^2\square{\bB\times\bb}_z\\
\mathscr{A}^s(\bQ,\bk)&=s(s^2-\al^2B^2),
\end{align}
where $\f_b=\tan^{-1}(b_y/b_x)$ and similarly for $\f_B$. Note that both
$\mathscr{A}_{1}^t(\bQ,\bk)$ and $\mathscr{A}_{0}^t(\bQ,\bk)$ are odd functions of the relative momentum $\bk$,
while $\mathscr{A}^s(\bQ,\bk)$ is even. This ensures that the wave function $\x(\bQ,\bk)$ is anti-symmetric under
interchange of two particles: $\bQ\rightarrow\bQ$, $\bk\rightarrow-\bk$ and $\s_1\leftrightarrow\s_2$.
(\ref{an2}) together with the Green function (\ref{GFapp}) represent the exact solution, and is the main
new technical result of this work.

The normalization constant $c_\bQ$ can be obtained from the orthonormality condition for the energy eigenstates $\ket{\psi(\bQ)}$, i.e. 
$\langle\psi(\bQ)|\psi(\bQ)\rangle=\langle\bQ|\bQ\rangle$ for all $\bQ$ (and $n$),
\beq
|c_\bQ|^2=
\square{V_0^2\int\frac{d^2\bk}{(2\p)^2}\bra{00}\cG^\dag(\bQ,\bk)\cG(\bQ,\bk)\ket{00}}^{-1}.
\eeq
The solution is complete once the energy spectrum $E_n(\bQ)$ is obtained for all the eigenstates.

\section{Properties of the bound molecular state}
\label{molstate}
If an energy eigenstate $(\bQ,n)$ describes a bound state of our two fermion system its spectrum
must satisfy $E_{n}(\bQ)<2E_{\rm min}$ for some values of $\bQ$. Here, $E_{\rm min}$ is the minimum value
in the single-fermion spectrum for Hamiltonian (\ref{HRD}), which is
\beq
\label{SPSpec}
E_\bk=\frac{k^2}{2m}\pm\al k\sqrt{1+\g^2+2\g\sin(2\f_k)},
\eeq
where $\f_k=\tan^{-1}(k_y/k_x)$. We find $E_{\rm min}=-m\al^2(1+\g)^2/2$. $\g=\be/\al$, again, is the relative
strength between the Rashba and Dresselhaus couplings.
We may now define the momentum-dependent binding energy $\De_{n}(\bQ)$ through $E_{n}(\bQ)
=2E_{\rm min}(1+\De_{n}(\bQ)/m\al^2)$. The condition for a bound state then translates to $\De_{n}(\bQ)>0$
for some $\bQ$.

\subsection{The binding energy at $\bQ=0$}
\label{be}
\begin{figure}[t]
\centering
\includegraphics*[scale=0.25]{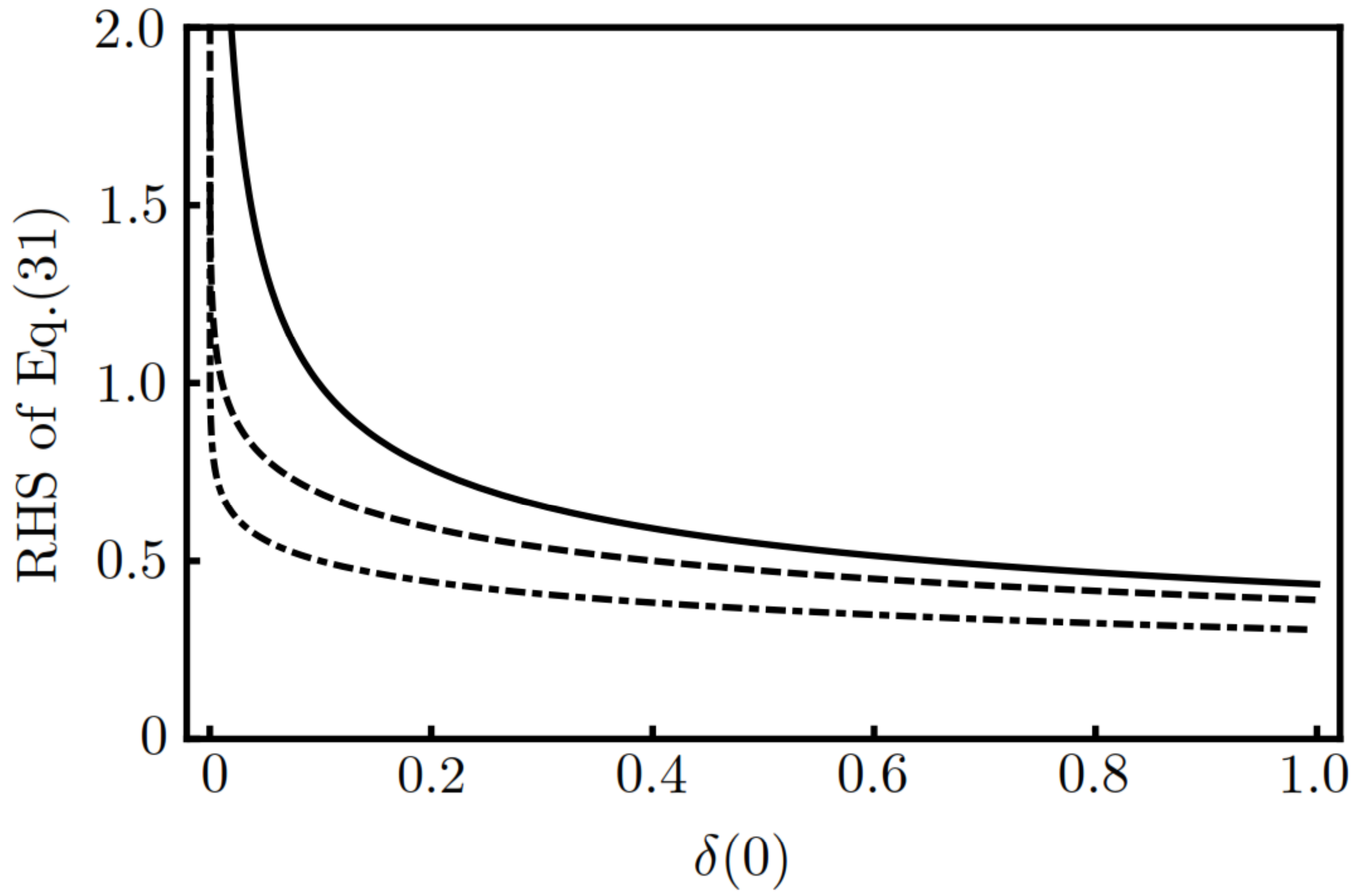}
\caption{\label{fig:SC} Plot of the right hand side of (\ref{SCQ02}) as a function of the binding energy
$\de(0)>0$. Here, we have chosen $\g=0$ for the solid line, $\g=0.1$ for the dashed line and $\g=0.5$ 
for the dot-dashed line. While an algebraic divergence is observed as $\de(0)\rightarrow 0$ in the
$\g=0$ case, a logarithmic divergence is observed in the presence of the Dresselhaus coupling ($\g>0$).
The UV cutoff is $\la=100$.}
\end{figure}

Owing to the oddness of the triplet coefficients in (\ref{an2}) with respect to $\bk$, one finds that only the singlet component 
survives once one integrates over $\bk$ in (\ref{cnQ}). This then leads to the eigenvalue equation for the bound state
\beq
\label{SCp}
\int\frac{d^2\bk}{(2\p)^2}\x(\bQ,\bk)=c_\bQ\ket{00}.
\eeq
Making use of (\ref{an}) leads to the self-consistency condition
\beq
-\frac{1}{V_0}=\int\frac{d^2\bk}{(2\p)^2}\bra{00}\cG(\bQ,\bk)\ket{00},
\eeq
which, together with (\ref{HRD}), (\ref{H0}) and (\ref{GFapp}), has an explicit form
\beq
\label{SC2D_gen}
\int\frac{d^2\bk}{(2\p)^2}\frac{s(s^2-\al^2B^2)}
{d(\bQ,\bk)}=-\frac{1}{V_0},
\eeq
Introducing the dimensionless variables,
\beq
\bq=\frac{\bQ}{m\al},\quad\bka=\frac{\bk}{m\al},\quad\de_n=\frac{\De_n}{m\al^2},\quad v_0=mV_0,
\eeq
and the dimensionless energy variable $\xi=\ka^2$ the self-consistency condition for $\bq=0$ reduces to
\beq
\label{SCQ02}
-\frac{2}{v_0}=\int_{0}^{\la}\frac{d^2\xi d\f}{(2\p)^2}\frac{\xi-e_n(0)}{(\xi-e_n(0))^2-4a(\f)\xi}.
\eeq
Here, we have defined the dimensionless energy spectrum $e_n(\bq)=e_t(1+\de_n(\bq))$, where 
$e_t=2e_{\rm min}=-(1+\g)^2$ is the dimensionless threshold energy and $a(\f)=1+\g^2+2\g\sin2\f$. 
We have also introduced the UV cutoff $\la\sim1/(m\al R_e)^2$ which is set by the characteristic radius of the interaction potential $R_e$. 
The typical $\de_n(0)$-dependence of the right hand side of (\ref{SCQ02}) is plotted 
in Fig. \ref{fig:SC} for $\g=0$, $\g=0.1$ and $\g=0.5$. We find that for any given attractive interaction 
$v_0<0$, there is a \textit{single} bound state. We label this state by $n=0$, but refrain from
explicitly writing the index. For the isotropic case 
(i.e. $\g=0$) and for $0<\de(0)\ll1$, we find that the binding energy is given by 
\beq
\label{BEQ0r}
\de(0)\approx\frac{|v_0|^2}{16}=:\de^R(0).
\eeq
In the weakly anisotropic regime (i.e. $0<\g/\de(0)\ll1$) we find the lowest $\g$-corrections to be
\beq
\de(0)\approx\de^R(0)\square{1-2\frac{\g}{\de^R(0)}+\frac{3}{2}\round{\frac{\g}{\de^R(0)}}^2},
\eeq
while in the strongly anisotropic regime (i.e. $\g/\de(0)\gg1$) we obtain
\beq
\label{BEQ0rd}
\de(0)\approx\frac{\la}{(1+\g)^2}e^{-\frac{8\p\sqrt{\g}}{|v_0|(1+\g)}}.
\eeq
Here, we have assumed $\la\gg1+\g$. We remind the reader that $\g=\be/\al$ is the relative
strength between the Rashba and Dresselhaus couplings. Similar calculations for the binding energy have been done
previously for various SO coupled Fermi gases.\cite{bcsbecsoc1,bcsbecsoc2,bcsbecsoc3,bcsbecsoc4,bcsbecsoc5}
However, the consideration of arbitrary two-dimensional Rashba and Dresselhaus coupling strengths in two spatial 
dimensions and the presentation of the exact wave function, to the best of our knowledge, has not been done.

The crossover in the binding energy from an algebraic dependence on
the interaction strength (c.f. (\ref{BEQ0r})) to exponential (c.f. (\ref{BEQ0rd})) is directly related to the crossover
in \textit{effective dimensionality} for bound state formation.\cite{socenhance} Let us say that a SO coupled system in $D$ spatial
dimensions has a set of single-particle minimum-energy states in momentum space that forms a $d$-dimensional
surface. The effective dimensionality for bound state formation is given by $D_{\rm eff}=D-d$. As can
be seen from the single-particle spectrum (\ref{SPSpec}) the isotropic case possesses a 1D manifold of minimum
states ($d=1$) while once the Dresselhaus coupling is finite, this manifold of minimum states is reduced to 
two points in momentum space ($d=0$). The effective dimensionalities for the two cases are then $D_{\rm eff}=1$ and
$D_{\rm eff}=2$, respectively. The exponentially small binding energy (\ref{BEQ0rd}) and the effective
dimension $D_{\rm eff}=2$ corresponding to that case is consistent with a bound state problem
in quantum mechanics of a particle moving in a potential well in two dimensions.

The binding energy can also be computed in the regime where $\de(0)\gg1$. Since $\g\sim1$, 
we will simply consider $\g=0$. If the cutoff scale is still the largest
scale, such that $\de(0)\ll\la$, we find the binding energy to be
\beq
\label{BEbig}
\de(0)\approx\la e^{-\frac{4\p}{|v_0|}}.
\eeq
\begin{figure}[t]
\centering
\includegraphics*[scale=0.36]{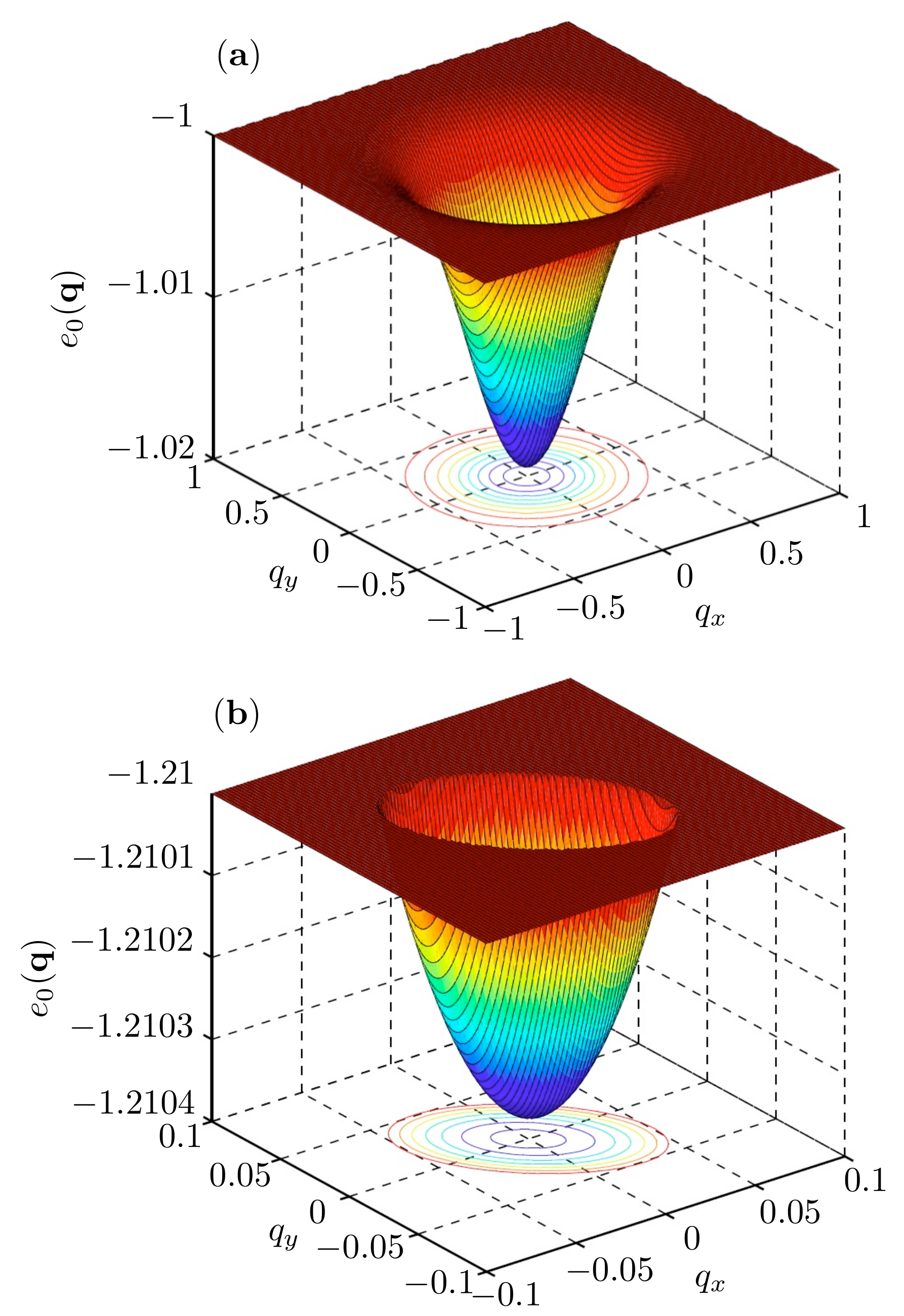}
\caption{\label{fig:disps} Numerical results for the dimensionless molecular spectrum $e(\bq)=E(\bQ)/m\al^2$ as a function
of the dimensionless molecular momentum $\bq$. (a) $\g=0$ and $|v_0|^{-1}=2$ are used. (b) $\g=0.1$
and $|v_0|^{-1}=1.5$ are used. Anistropy in the molecular effective mass can be observed in (b) for finite
Dresselhaus coupling (c.f. (\ref{BEQfrd})). The plot shows the bound state ceases to exist for momenta larger
than the critical momentum $q_c\sim\sqrt{\de(0)}$.}
\end{figure}

\subsection{The molecular spectrum}
\label{spectrum}
We first provide an approximate analytic
expression for the spectrum valid for $q^2/\de(0)\ll1$, from which the effective mass of the molecule can be extracted.
In the weakly isotropic limit, the binding energy to $\mathscr{O}(q^2)$ reads
\beq
\label{BEQfr}
\de(\bq)\approx\de(0)-\frac{q^2}{8}\square{1+\frac{3}{2}\frac{\g}{\de^R(0)}\sin(2\f_q)},
\eeq
where $\f_q=\tan^{-1}(q_y/q_x)$. For $\g=0$, the isotropy in the spectrum is restored as expected.  
The corresponding effective mass for the molecule is then given by
\beq
M_{\rm eff}(\g)\approx4m\square{1-\frac{3}{2}\frac{\g}{\de^R(0)}\sin(2\f_q)}.
\eeq
In the strongly anisotropic regime, the binding energy to $\mathscr{O}(q^2)$ reads
\beq
\label{BEQfrd}
\de(\bq)\approx\de(0)-\frac{q^2}{8}\frac{1+6\g+\g^2+(1-\g)^2\sin(2\f_q)}{(1+\g)^4},
\eeq
with the corresponding effective mass of
\beq
M_{\rm eff}(\g)\approx\frac{4m(1+\g)^2}{1+6\g+\g^2+(1-\g)^2\sin(2\f_q)}.
\eeq
Interestingly, we find that isotropy in the
effective mass is restored for equal Rashba and Dresselhaus interaction strengths ($\g=1$), the so-called persistent
spin helix point\cite{PSH1,PSH2,PSH3}.
Let us now consider the effective mass of a free fermion, $m_{\rm eff}(\g)$,  near the bottom of the single-particle 
spectrum (\ref{SPSpec}). We find that the \textit{molecular} effective mass and the \textit{single-fermion} effective mass 
are related in a one-to-one relation, namely, $M_{\rm eff}(\g)=2m_{\rm eff}(\g)$. 
The dimensionless molecular spectrum $e(\bq)=E(\bQ)/m\al^2$ can be straightforwardly obtained through the relation $e(\bq)=e_t(1+\de(\bq))$.

The molecular spectrum can be obtained numerically for arbitrary values of $\bq$. The results for the spectrum 
as a function of $\bq$ are plotted in Fig. \ref{fig:disps} for $\g=0$ and $\g=0.1$, respectively. $|v_0|$ was chosen to be
$\mathscr{O}(1)$ such that we are in the regime of $\de(0)\ll1$. We see the anisotropy in the spectrum
for $\g=0.1$ (Fig. \ref{fig:disps}(b)). We also see that for large enough momenta, $q>q_c$, a bound state
ceases to exist. The scale for the critical momentum is set by $q_c\sim\sqrt{\de(0)}$. The existence of this critical momentum 
tells us that the fermions do not remain bound once the molecular kinetic energy exceeds the binding energy.

An estimate of the binding energy can be made if a gas of these fermionic molecules is rotated. We consider the gas confined 
to the $xy$-plane and a rotation about the $z$-axis. For simplicity, we will take the pure Rashba case here, i.e. $\g=0$. 
In the dilute limit, where the gas can be treated as a classical 
(Maxwell-Boltzmann) gas, the velocity field under the rotation $\bm{\W}=\W\mathbf{z}$ is given by a
rigid rotation $\mathbf{v}(R)=\W R\bm{\phi}$, where $\bm{\phi}$ is the unit tangent vector on the $xy$-plane and
$R=\sqrt{x^2+y^2}$ is the radial distance from the axis of rotation. The critical momentum scale then introduces a 
critical distance scale $R_c$, where, for $R<R_c$, the molecules are still bound but, for $R>R_c$, we have a gas
of unbound fermions. If we introduce a dimensionless distance $r=R\W/\al$ the critical distance is given by 
$r_c=q_c/4\sim\sqrt{\de(0)}$. Therefore, the radial distance at which the two phases of bound and unbound
fermions meet gives an estimate of the binding energy.

\subsection{Effects due to a Zeeman field}
\label{zeeman}
We briefly discuss the effects of a synthetic Zeeman field, ${\bf H}=(H_x,H_y,H_z)$, which couples to the pseudo-spin
of the fermions. We introduce a Zeeman term, $\hH_Z=-(H_x\hpm_x
+H_y\hpm_y+H_z\hpm_z)$ to the single-particle Hamiltonian (\ref{HRD}). In Appendix \ref{zeemanapp} the self-consistency 
condition is rederived for a Zeeman field oriented in a general direction in $\mathbb{R}^3$. Here, we will explicitly consider the field pointing in the 
$z$-direction, and for the pure Rashba case where $\g=0$. We find that the self-consistency condition at $\bq=0$
is given by
\beq
\label{SCQ0h}
-\frac{2}{v_0}=\int_0^{\la}\frac{d^2\xi}{2\p}\frac{(\xi-e_0(0))^2-4h^2}
{(\xi-e_0(0))[(\xi-e_0(0))^2-4\xi-4h^2]},
\eeq
where the dimensionless Zeeman field $h=H/m\al^2$.
Since the threshold energy for molecular formation is now $e_t=-(1+h^2)$, the dimensionless 
energy for the bound state at $\bq=0$ reads $e_0(0)=-(1+h^2)(1+\de(0))$. For $h<h_c=1$, (\ref{SCQ0h}) 
gives
\beq
-\frac{4}{v_0}\approx\frac{1-h^2}{\sqrt{1+h^2}}\frac{1}{\sqrt{\de(0)}},
\eeq
where again the approximation holds for $\de(0)\ll1$ and $\la\gg1$. 
For $h<h_c$, a bound state exists for all $|v_0|$, but the binding energy approaches zero as $h\rightarrow h_c$.
Once $h>h_c$, the integral on the right hand side of (\ref{SCQ0h}) becomes bounded as $\de(0)\rightarrow0$ and, therefore, 
a bound state ceases to exist for sufficiently weak attractive interaction strengths. This introduces a quantum phase transition
at a critical coupling $v^c_0$ separating phases with bound and unbound fermions.

\section{Low-density Rashba-Dresselhaus molecular gas}
\label{probe}
Let us now consider a low-density gas of $N$ tightly-bound Rashba-Dresselhaus molecules confined to two spatial dimensions. 
We assume that the gas is equilibrated at some temperature $T$, which satisfies $T_{BKT}\ll T\ll\De$. Here, $T_{BKT}$ is the temperature
at which the gas undergoes a Berezinskii-Kosterlitz-Thouless phase transition into a superfluid, and $\De$
is the molecular binding energy. Presumably, in this temperature range, the gas is in the dilute limit, $n\la_T^2\ll1$, 
where $n$ is the areal density of the molecules and  $\la_T=h/\sqrt{2\p\mathscr{M}k_BT}$, with $\mathscr{M}:=\mbox{min}\{M_{\rm eff}(\g)\}$, 
is the mean thermal wavelength evaluated for the smallest molecular effective mass. The gas may then be modeled as a thermal 
Boltzmann gas of uncondensed molecules. 

A molecule can interact with other atoms and molecules in the gas. Indeed, scattering between atoms and bound molecules
as well as between two molecules was considered in depth in many works.\cite{int1,int2,int3,int4,int5} Here, we assume that the gas
is sufficiently dilute so that we may neglect atom-molecule and molecule-molecule interactions to first order. 

Our aim first is to obtain TOF velocity and spin distributions for a single fermion atom for the molecular gas. These distributions 
can be inferred from a spin-resolved TOF experiment, where both the trap and the SO coupling are turned off and the \textit{depaired}
fermions are allowed to expand freely.
We then contrast these distributions to corresponding 
distributions for a Boltzmann gas of unbound (non-interacting) fermionic atoms and discuss the striking differences between the two
cases.

\subsection{Qualitative discussion of the result}
\label{qualitative}
For the gas of non-interacting molecules (``gas $A$") single-fermion distributions must be extracted from the \textit{single-molecule} density
matrix. For the gas of non-interacting fermion atoms (``gas $B$") the corresponding distributions are obtained directly from the 
\textit{single-fermion} density matrix. For gas $A$ the single-molecule density matrix is given by 
\beq
\label{dmgasa}
\dm_A=\frac{e^{-\be\cH}}{\tr\{e^{-\be\cH}\}},
\eeq
where the Hamiltonian $\cH=\cH_0+\cH_{\rm int}$ was given in (\ref{H0}) and (\ref{Hint}) in Sec. \ref{exactsol}.
For gas $B$, the single-fermion density matrix reads
\beq
\dm_B=\frac{e^{-\be\hH_{RD}}}{\tr\{e^{-\be\hH_{RD}}\}},
\eeq
where the Hamiltonian was given in (\ref{HRD}).
The key difference in the single-fermion distributions for gases $A$ and $B$ stems from the fact that while the single-fermion 
momentum eigenstate for gas $B$, $\ket{\bk}$, is an eigenstate of $\hH_{RD}$, the two-fermion momentum eigenstate, 
$\ket{\bQ,\bk}$, for gas $A$ is not an eigenstate of $\cH$. For gas $B$ the momentum operator commutes with the Hamiltonian, 
and the velocity distribution for the Boltzmann
gas is trivially given by
\beq
\label{P1}
P(\bk)=\bra{\bk}\dm_B\ket{\bk}\propto e^{-\be E_\bk},
\eeq
where $E_\bk$ was given in (\ref{SPSpec}). For gas $A$, we first obtain the diagonal elements of the molecular density matrix,
$\bra{\bQ,\bk}\dm_A\ket{\bQ,\bk}$, and the velocity distribution is extracted by tracing out one of the fermions. Since $\ket{\bQ,\bk}$
is not an eigenstate of $\cH$, we must introduce the energy eigenstates $\ket{\psi_n(\bQ)}$ in order to replace $\cH$ by its 
expectation value. This results in the diagonal elements which schematically has the form
\begin{align}
\label{P12}
P(\bk_1,\bk_2)&=\bra{\bQ,\bk}\dm_A\ket{\bQ,\bk}\nn\\
&\propto \x^\dag(\bQ,\bk)\x(\bQ,\bk)e^{-\be E(\bQ)},
\end{align}
where $E(\bQ)$ is the molecular spectrum obtained in Sec. \ref{spectrum} and $\x(\bQ,\bk)$ is the spinor wave function
obtained in (\ref{an2}). The extra factor involving $\x$ in (\ref{P12}) indicates 
that each fermion is correlated with its partner fermion due to interactions. In the free fermion case (c.f. (\ref{P1})),
such correlations are clearly absent. 

The correlations between fermions in gas $A$ has an imprint on the various momentum distributions 
and make them distinct from the corresponding distributions in gas $B$. Moreover, these distinctions can be
made within first-moment expectation values (e.g. $\ang{n(\bk)}$).
These differences can, in principle, be inferred from TOF 
experiments. For small binding energies (i.e. $\de(0)\ll1$)  the \textit{qualitative} shapes 
of the various momentum distributions, in general, appear similar for both gases. The key difference arises 
in the width of the peak features found in the distributions. For free fermions (gas $B$) these peak features should have a gaussian 
profile with the width set by temperature, as it is clear from
(\ref{P1}). In contrast, the peak features for gas $A$ has a square-Lorentzian profile with the width set by the 
binding energy scale. In fact, for gas $A$, the peak features are determined by the
wave function $\x(\bQ,\bk)$, and not the Boltzmann factor, which primarily determines the centre-of-mass 
distribution of the molecules (c.f. (\ref{P12})).

For $\tau:=k_BT/m\al^2\ll\de(0)$, the width in the peak structures for gas $A$ should be markedly broader
than those for gas $B$. The peak widths for gas $B$ can be adjusted by changing the temperature while such variation 
should not occur for gas $A$. On the other hand, increasing the attractive interaction strength, thus increasing the
binding energy, while holding the temperature fixed should lead to a broadening in the width of the peaks for gas $A$ only.

\subsection{Quantitative results for various distributions}
\label{quantitative}
We now provide detailed calculations of the single-fermion distributions for gas $A$ described above. 
Recall that the (dimensionless) molecular dispersion was given by $e(\bq)=e_t(1+\de(\bq))$, 
with $\bq$-dependent binding energies (\ref{BEQfr}) or (\ref{BEQfrd}) which are valid for $q^2\ll\de(0)$. The validity 
of (\ref{BEQfr}) or (\ref{BEQfrd}) is ensured in our temperature regime, since $\tau=k_BT/m\al^2\ll\de(0)$, such that essentially 
all of the fermions are bound and occupy single-molecule states very close to $\bq=0$. The diagonal elements of the single-molecule 
density matrix (\ref{dmgasa}) can then be approximately written as (see Appendix \ref{P12details})
\begin{multline}
\label{P12exp}
P(\bka_1,\bka_2)\approx\\
\frac{|c_\bq|^2\bra{00}\cG^\dag(\bq,\bka)\cG(\bq,\bka)\ket{00}e^{-e(\bq)/\tau}}
{\int_{\bq,\bka} |c_\bq|^2\bra{00}\cG^\dag(\bq,\bka)\cG(\bq,\bka)\ket{00}e^{-e(\bq)/\tau}},
\end{multline}
where $\int_{\bq,\bka}=\int\frac{d^2\bq}{(2\p)^2}\frac{d^2\bka}{(2\p)^2}$. Here again, $\bq$ and $\bka$ are dimensionless
momenta and $\bq=\bka_1+\bka_2$ and $\bka=(\bka_1-\bka_2)/2$.
\begin{figure}[t]
\centering
\includegraphics*[scale=0.45]{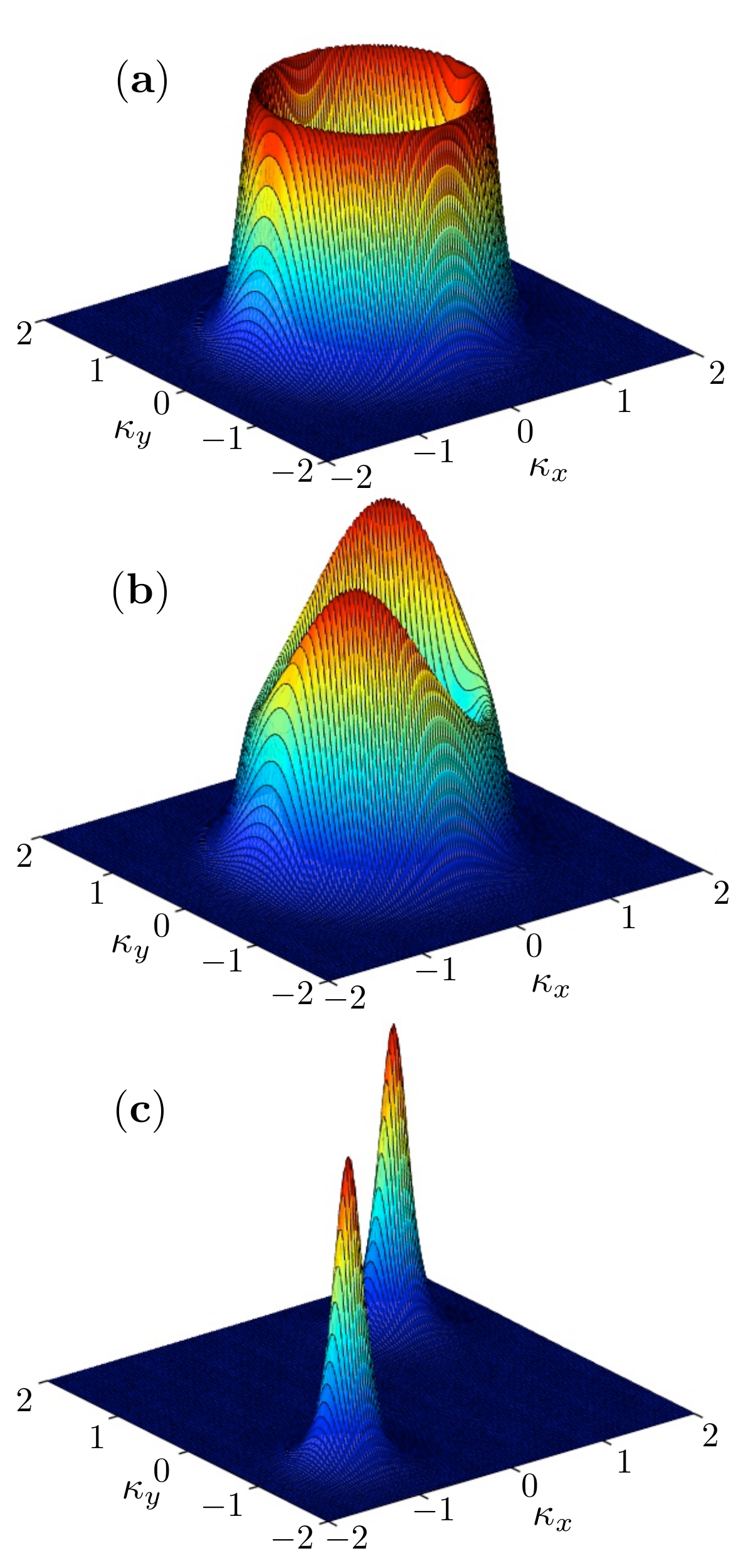}
\caption{\label{fig:VDs} Single-fermion velocity distribution for the molecular gas for three different relative SO interaction strengths:
(a) $\g=0$; (b) $\g=0.005$; and (c) $\g=0.5$. Binding energy $\de(0)=0.05$ is used, and the temperature is
taken to be small such that $\tau\ll\de(0)$.}
\end{figure}
The velocity distribution for a single fermion can be obtained by integrating out the other,
\beq
\label{veldist}
P(\bka_1)=\int\frac{d^2\bka_2}{(2\p)^2}P(\bka_1,\bka_2).
\eeq
Note that $P(\bka_1,\bka_2)=P(\bka_2,\bka_1)$ so one could have integrated out either one or the other
electron and would have arrived at the same probability. 

The single-fermion spin distributions can also be obtained analogously. 
The three spin operators for a fermion are $\hat S_i=\hpm_i/2$, where $i=x,y,z$. The $i$-th component of the 
spin density is then given by
\begin{multline}
\label{spindensity}
S_i(\bka_1):=\ang{\hat S_i}(\bka_1)=\int\frac{d^2\bka_2}{(2\p)^2}\\
\times\frac{|c_\bq|^2\bra{00}\cG^\dag(\bq,\bka)[\hat S_i\otimes\hI]\cG(\bq,\bka)\ket{00}e^{-e(\bq)/\tau}}
{\int_{\bq,\bka} |c_\bq|^2\bra{00}\cG^\dag(\bq,\bka)\cG(\bq,\bka)\ket{00}e^{-e(\bq)/\tau}},
\end{multline}
Note here that the matrix $\hat S_i\otimes\hI$ acts in spin space spanned by the basis $\ket{\s_1,\s_2}=\{\ket{\up\up},\ket{\up\down},
\ket{\down\up},\ket{\down\down}\}$. We find that $S_z(\bka_1)=0$, which is expected from solving a single-fermion quantum 
mechanical problem with Hamiltonian (\ref{HRD}).

At low temperatures but still well above $T_{BKT}$, 
the Boltzmann factor in (\ref{veldist}) and (\ref{spindensity}) is a strongly peaked function at 
$\bq=0$ with contributions becoming exponentially small for $q>\sqrt{\tau}$. In the low temperature regime 
we are considering ($\tau\ll\de(0)$) the remaining factors in (\ref{veldist}) and (\ref{spindensity}) give a slowly-varying 
function of $\bq$ on the scale of $\sqrt{\tau}$. We may then drop the $\bq$-dependence in those factors and replace it by 
0. As a result, the integral over $\bka_2$ can be done trivially and we arrive at
\beq
\label{veldistapp}
P(\bka)\approx C\bra{00}\cG^\dag(0,\bka)\cG(0,\bka)\ket{00},
\eeq
where $C$ is a normalization constant.
Similarly, the spin densities can be approximately written as
\beq
\label{spindensityapp}
S_i(\bka)\approx C\bra{00}\cG^\dag(0,\bka)[\hat S_i\otimes\mathbb{I}]\cG(0,\bka)\ket{00}.
\eeq
\begin{figure}[t]
\centering
\includegraphics*[scale=0.45]{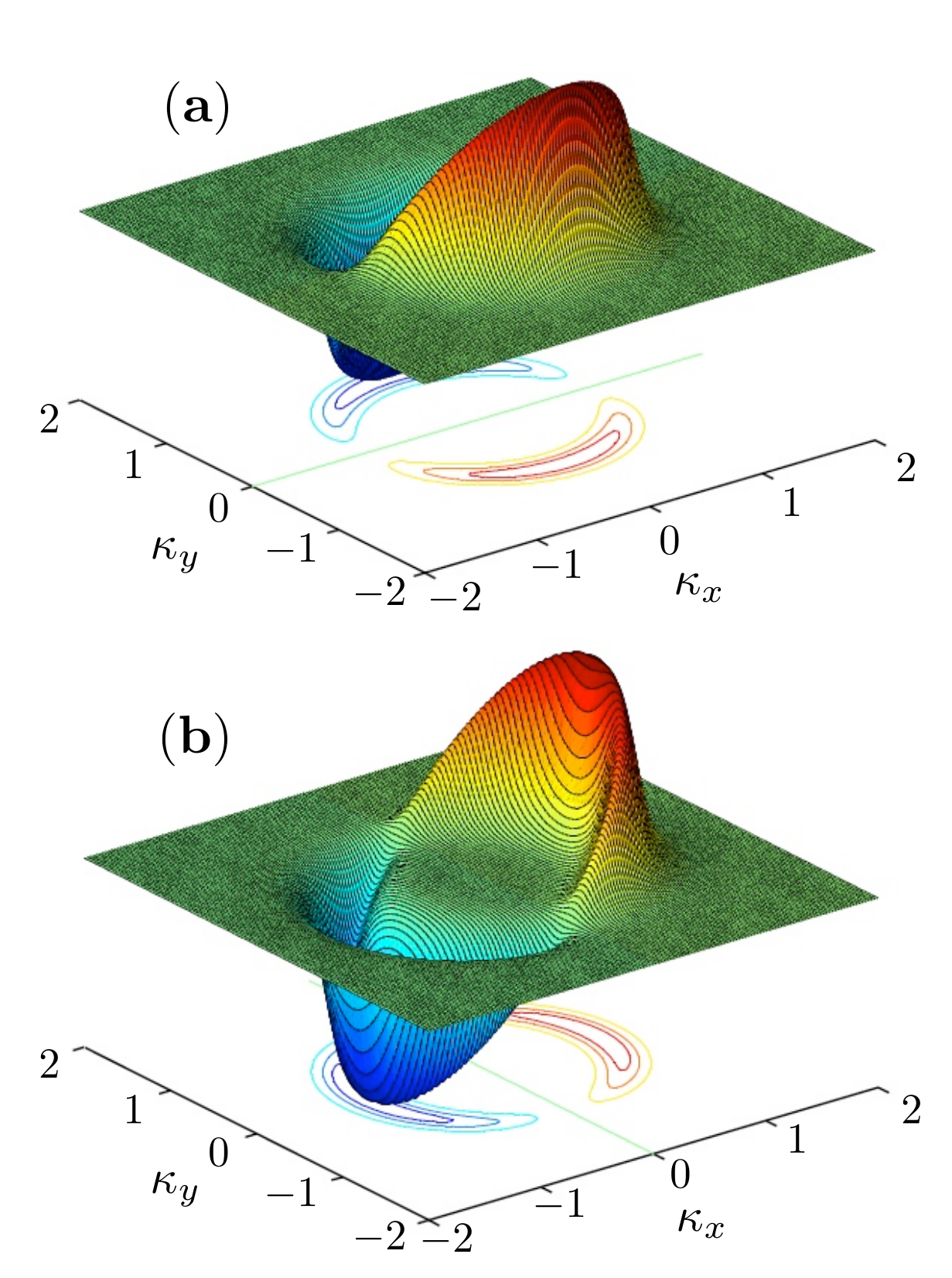}
\caption{\label{fig:S_R} Single-fermion spin densities in momentum space for $\g=0$. 
$S_x(\bka)$ and $S_y(\bka)$ are plotted in (a) and (b), respectively. Binding energy $\de(0)=0.05$ is used, and the 
temperature is taken to be small such that $\tau\ll\de(0)$.}
\end{figure}

\subsection{Velocity distribution}
\label{sec:veldist}
Plots of $P(\bka)$ are shown in Fig. \ref{fig:VDs} for three different values of $\g$: (a) $\g=0$; (b) $\g=0.005$;
and (c) $\g=0.5$. The $\bq=0$ binding energy was taken to be $\de(0)=0.05$ and $\tau=10^{-6}$. We find that the maxima in 
the velocity distributions occur for values of momenta where the minima in the single-particle spectrum (\ref{SPSpec})
occur. For $\g=0$ the spectrum  has a ring of degenerate minima at $\ka=1$, and this is reflected in the distribution in 
Fig. \ref{fig:VDs}(a).
The Dresselhaus interaction breaks this degeneracy and the spectrum yields two minima at $(\ka_x,\ka_y)
=\pm(1+\g,1+\g)/\sqrt{2}$. The two peaks in Fig. \ref{fig:VDs}(c) coincide again with the locations of these minima.
The distribution in the crossover regime between the isotropic and strongly anisotropic limits is plotted in
Fig. \ref{fig:VDs}(b). 
\begin{figure}[t]
\centering
\includegraphics*[scale=0.45]{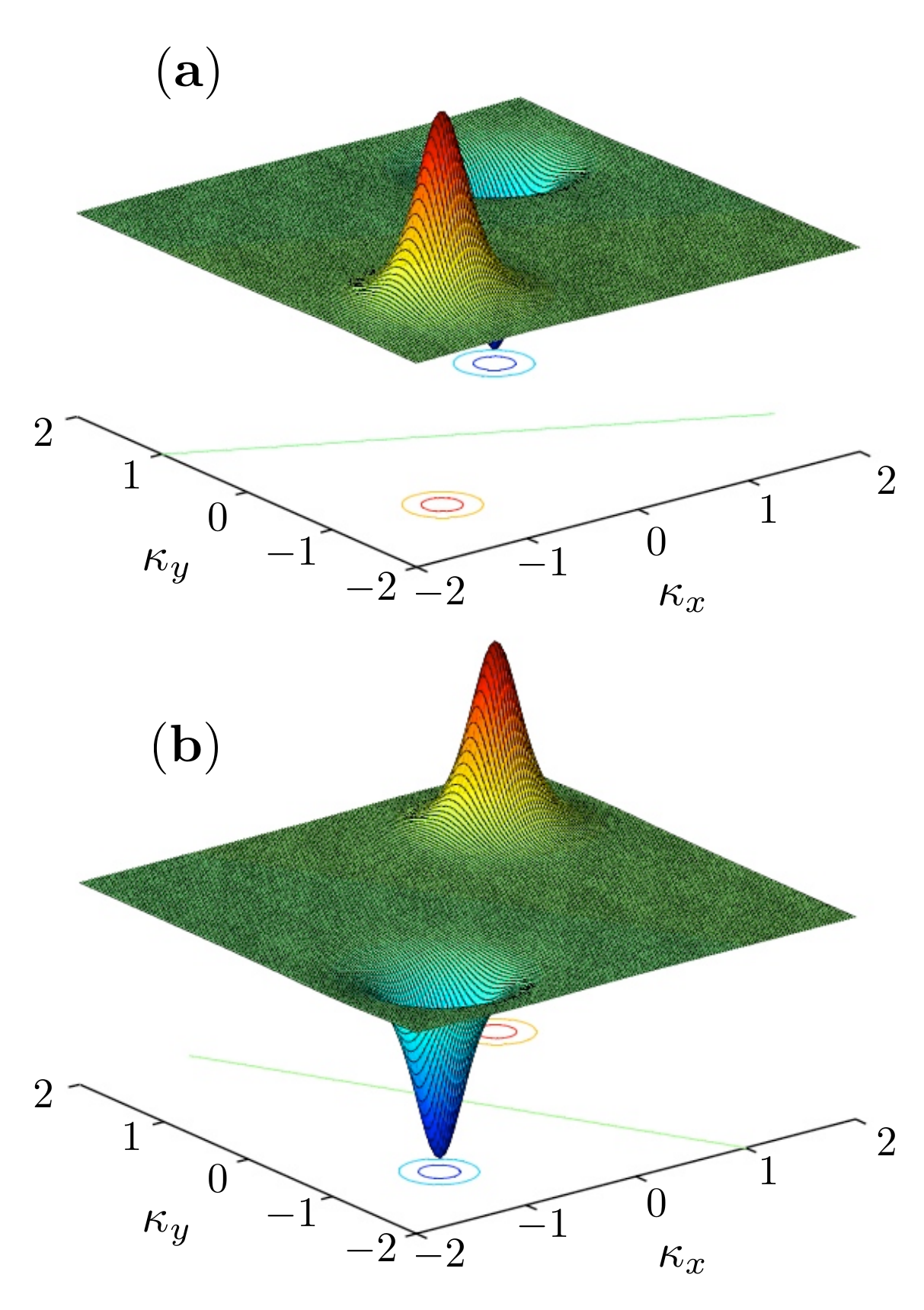}
\caption{\label{fig:S_RD} Single-fermion spin densities in momentum space for $\g=0.5$. 
$S_x(\bka)$ and $S_y(\bka)$ are plotted in (a) and (b), respectively. Binding energy $\de(0)=0.05$ is used, and the 
temperature is taken to be small such that $\tau\ll\de(0)$.}
\end{figure}

\begin{figure}[t]
\centering
\includegraphics*[scale=0.45]{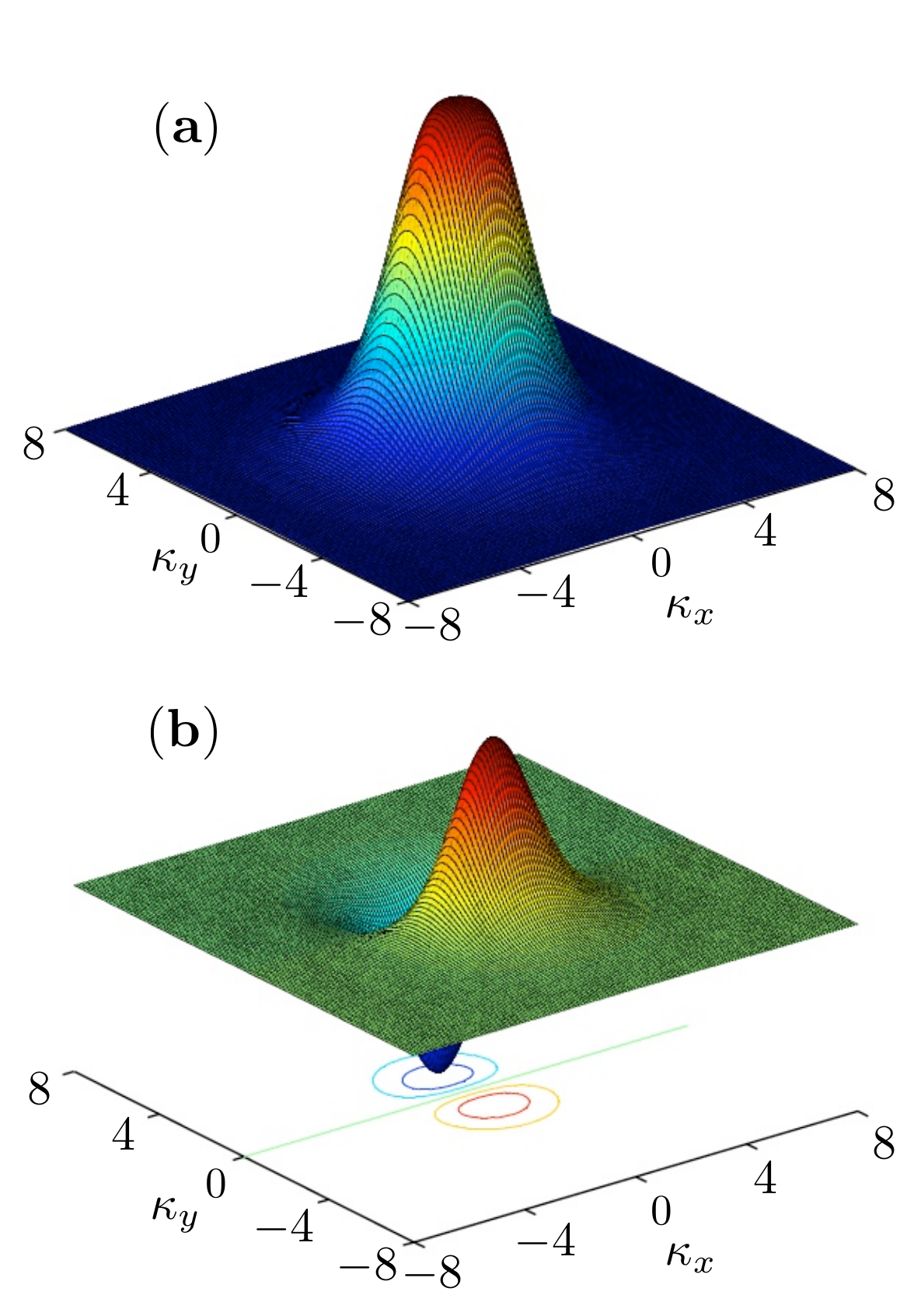}
\caption{\label{fig:bigBE} Single-fermion velocity and $S_x$ distributions with $\g=0$ 
in the large binding energy limit (i.e. $\de(0)\gg1$); here, we have taken $\de(0)=5$, and the temperature is 
still taken to be small such that $\tau\ll1$.}
\end{figure}

The fact that large weights are observed at spectrum minima is expected as we are in
the dilute and low temperature limits where most of the fermions are occupying momentum states near the band
minima. Although corresponding distributions for free fermions (gas $B$) will be qualitatively similar to the 
distributions obtained here for gas $A$, they are quantitatively different. For the pure Rashba case ($\g=0$), for instance, 
the width of the ring of maxima for gas $B$ is set by the temperature (c.f. (\ref{P1})). In particular, the peak profile has a 
gaussian profile with the width set by $\sqrt{\tau}$. In contrast, for gas $A$, a radial cut of the velocity distribution is 
essentially proportional to $\x^\dag(0,\bka)\x(0,\bka)$ and yields 
\begin{align}
P(\ka)&\propto\frac{(\ka^2+1+\de(0))^2+4\ka^2}{[(\ka+1)^2+\de(0)]^2[(\ka-1)^2+\de(0)]^2}\nn\\
&\approx\frac{(2+\de(0))^2+4}{[4+\de(0)]^2[(\ka-1)^2+\de(0)]^2}
\end{align}
for $\de(0)\ll1$. Therefore, the ring of maxima has a square-Lorentzian profile with the width set by $\sqrt{\de(0)}$. 
In the limit $\de(0)\gg\tau$, we would thus expect the peak width to be much broader for gas $A$ than the corresponding 
distribution for gas $B$. If $\de(0)$ is gradually increased while keeping the temperature fixed the ring of maxima for 
the molecular gas will get progressively broader. Once $\de(0)\gg 1$, the ring can no longer be resolved and one
obtains a single broad peak centred at $\bka=0$. This is illustrated in Fig. \ref{fig:bigBE}(a) 
where we have evaluated (\ref{veldistapp}) for $\de(0)=5$ (c.f. (\ref{BEbig})). This evolution of the velocity 
distribution as $\de(0)$ is increased is unique to that molecular gas, for, in the case of free fermions, the ring of 
maxima should remain sharp with the width set by $\sqrt{\tau}$.

\subsection{Spin density distributions}
\label{sec:spindensity}
Spin densities in momentum space, $S_x(\bka)$ and $S_y(\bka)$, are plotted in Figs. \ref{fig:S_R} and \ref{fig:S_RD} for the 
regime $\tau\ll\de(0)\ll1$. We consider $\g=0$ and $\g=0.5$ in Figs. \ref{fig:S_R} and \ref{fig:S_RD}, respectively.  In analogy 
to the velocity distribution, the spin-densities obtained here appear very similar to the corresponding spin-densities obtains by 
solving a single-fermion quantum mechanical problem with Hamiltonian (\ref{HRD}). However, again as in the case of the velocity
distribution, the widths of the maxima are set by the binding energy scale. In Fig. \ref{fig:bigBE}(b), $S_x$ density for the pure 
Rashba case is plotted for $\de(0)=5$. There, the half-rings of maxima and minima, as seen in Fig. \ref{fig:S_R}(a), are no 
longer resolved due to the large binding energy.

\subsection{Detecting the distributions in TOF experiments}
\label{experiments}
These momentum distributions will be directly observable through a spin-resolved TOF measurement.\cite{nist1}
The TOF signature will be dependent on the SO scheme used. In what follows we assume the 
effective SO coupling is induced using the $N$-level scheme, with $N=4$~\cite{nlevel}, which can be implemented 
in the alkalis such as ${}^6 \textrm{Li}$. In this scheme, the dressed states have the form 
\begin{equation}
\ket{D_a} = \frac{1}{2} \sum_{j=1}^4 e^{ i \pi aj / 2} \ket{\tilde{j}}
\end{equation}
where $\ket{\tilde{j}} = e^{i \KK_j \cdot \rr} \ket{j}$ is a bare hyperfine state, $\ket{j}$, boosted by $\KK_j =
m\al[- \sin(\pi j / 2) \mathbf{e}_x + \cos(\pi j / 2) \mathbf{e}_y]$. The two pseudo-spin states are given by 
$\ket{\uparrow} = \ket{D_1}$ and $\ket{\downarrow} = \ket{D_2}$. 

\begin{figure}[t]
\centering
\includegraphics*[scale=0.5]{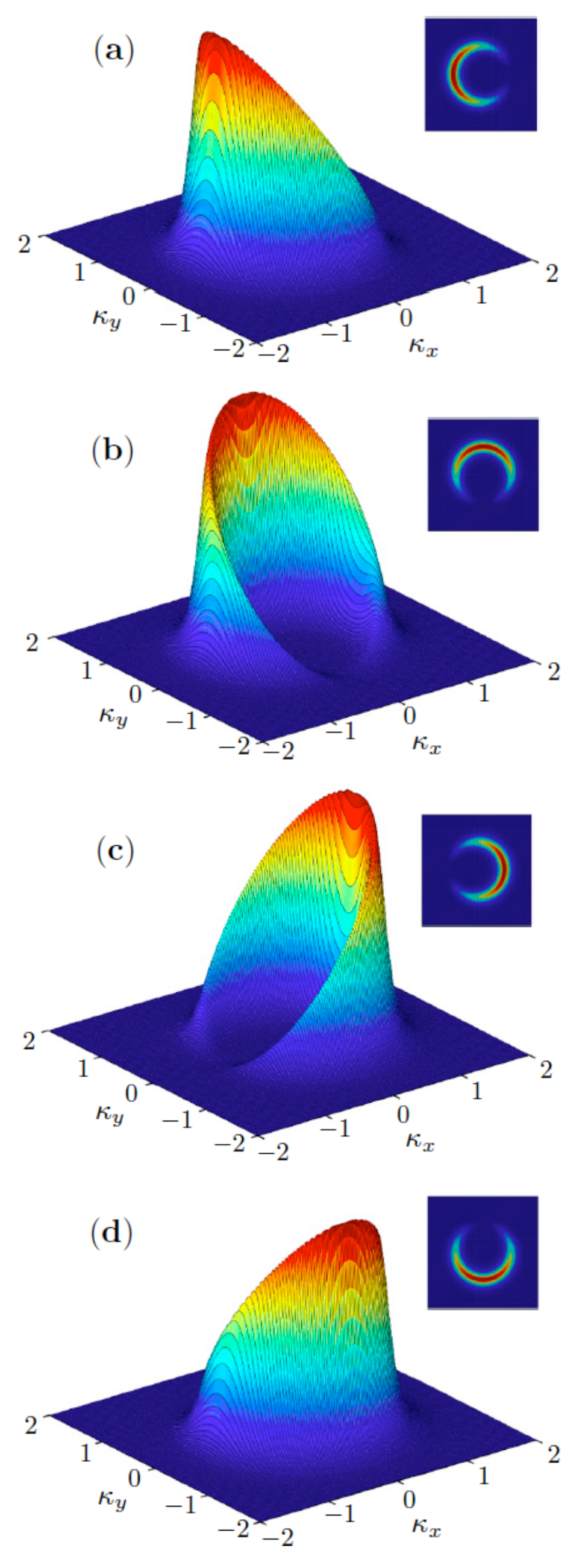}
\caption{\label{fig:BareHFS} Single-fermion velocity distribution for the four bare spin states $\ket{1}, \ket{2}, \ket{3}, \ket{4}$ are plotted in (a), (b), (c) and (d), respectively, for $\g=0$ in the small binding energy regime (i.e. $\de(0)\ll1$). Here, $\de(0)=0.05$ is used, and the temperature is taken to
be small such that $\tau\ll\de(0)$.}
\end{figure}
The velocity-spin distributions shown in Fig. \ref{fig:S_R} and \ref{fig:S_RD} can be inferred through a spin-resolved TOF measurement. 
Such a measurement will give the velocity distribution $\ev{ \ket{j} \bra{j} } = \ev{ \frac{1}{4} \sum_{a=1}^4 e^{- i \pi (a - a^\prime) j / 2} \ket{D_a}\bra{D_{a^\prime}} }$. At low temperatures the state $\ket{D_3}$ and $\ket{D_4}$ will not be populated, and the expectation value reduces to $\ev{\ket{j}\bra{j}} = \frac{1}{4} \ev{ \ket{\uparrow}\bra{\uparrow} + \ket{\downarrow}\bra{\downarrow} + e^{ i \pi j / 2} \ket{\uparrow}\bra{\downarrow} +  e^{ - i \pi j / 2} \ket{\downarrow}\bra{\uparrow} } $. These states have the equivalent momentum distributions
\begin{eqnarray}
\ev{\mathcal{P}_1} (\boldsymbol{\kappa}) & = & \frac{1}{4} (P(\boldsymbol{\kappa}) - 2 S_y(\boldsymbol{\kappa})) \\
\ev{\mathcal{P}_2} (\boldsymbol{\kappa}) & = & \frac{1}{4} (P(\boldsymbol{\kappa}) - 2 S_x(\boldsymbol{\kappa})) \\
\ev{\mathcal{P}_3} (\boldsymbol{\kappa}) & = & \frac{1}{4} (P(\boldsymbol{\kappa}) + 2 S_y(\boldsymbol{\kappa})) \\
\ev{\mathcal{P}_4} (\boldsymbol{\kappa}) & = & \frac{1}{4} (P(\boldsymbol{\kappa}) + 2 S_x(\boldsymbol{\kappa})) 
\end{eqnarray}
where $\mathcal{P}_j = \ket{j}\bra{j}$ is a projective measurement into the bare spin state $\ket{j}$. It is therefore possible to reconstruct the 
$P(\boldsymbol{\kappa})$,  $S_x(\boldsymbol{\kappa})$ and $S_y(\boldsymbol{\kappa})$ momentum distributions from a TOF measurement. 

\begin{figure}[t]
\centering
\includegraphics*[scale=0.5]{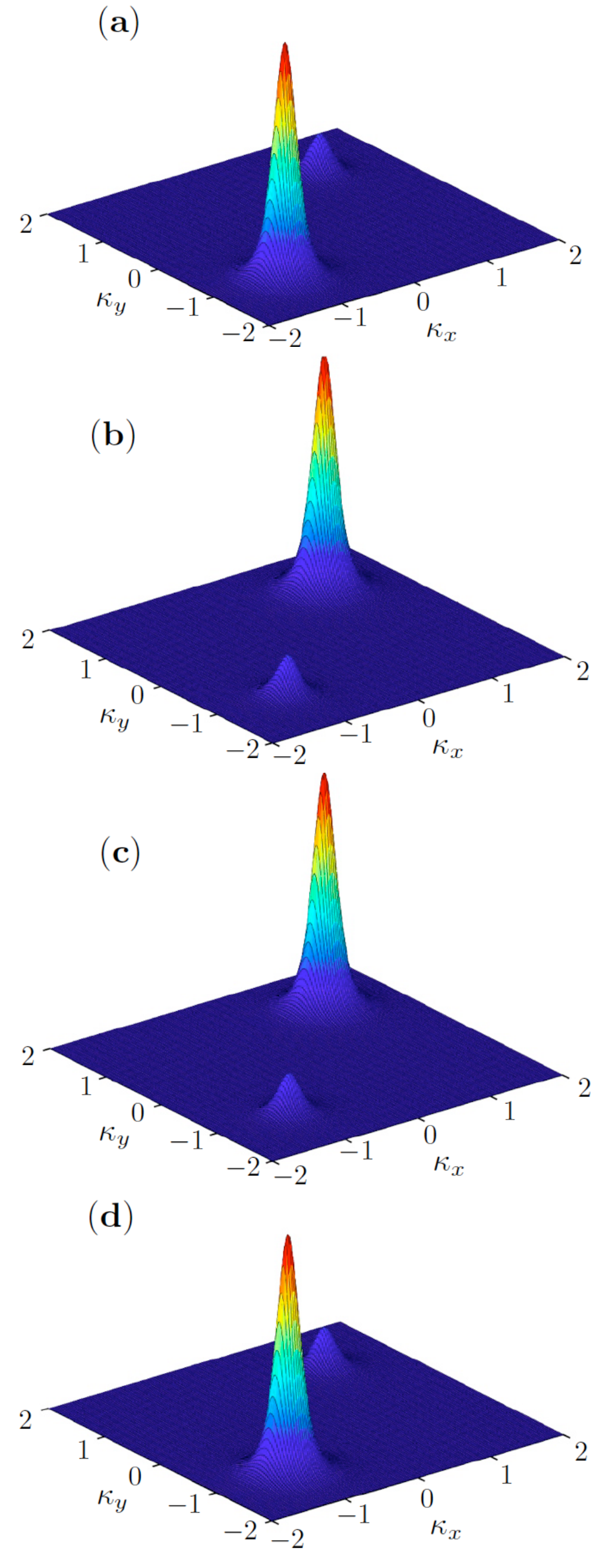}
\caption{\label{fig:BareHFSRD} Single-fermion velocity distribution for the four bare spin states $\ket{1}, \ket{2}, \ket{3}, \ket{4}$ are plotted in (a), (b), (c) and (d), respectively, for $\g=0.5$ in the small binding energy regime (i.e. $\de(0)\ll1$). Here, $\de(0)=0.05$ is used, and the temperature is taken to
be small such that $\tau\ll\de(0)$.}
\end{figure}
These spin distributions are plotted in Figs. \ref{fig:BareHFS} and \ref{fig:BareHFSRD}. In the pure Rashba limit, the momentum 
distribution of the four bare states have the same structure up to a $\pi/2$ rotation in momentum space. In the anisotropic limit, 
$\gamma \neq 0$, the bare spins have the same two-peak structure, but with different relative amplitudes. Similar to the 
momentum distribution, the bare spin distribution can be distinguished from the case of non-interacting SO coupled fermions 
by the dependence of the width of the distribution on the interaction strength.

\section{Conclusion}
\label{conc}
We investigate the properties of a low-density molecular gas of Rashba-Dresselhaus fermions in two spatial dimensions. 
The gas is considered at sufficiently high temperatures such that it can be considered as
a thermal gas of uncondensed tightly bound molecules. The description of the gas is based on the exact solution to a quantum 
mechanical problem of two Rashba-Dresselhaus fermions interacting via an attractive short-ranged $s$-wave interaction. 
We compute the single-fermion density matrix for the gas and evaluate the fermion velocity and spin distributions. By making
comparisons to corresponding distributions for a Boltzmann gas of free fermions we show that these various distributions
can be used to probe pairing of fermions in the molecular gas. Moreover, we find that the signatures of
pairing appear in first-moment expectation values. We discuss a spin-resolved TOF measurement from which the various
distributions can be inferred in an experiment. This result is not specific to SO coupled systems considered in
this work. Analogous signatures of correlations should appear generally in single-fermion distributions for a system composed of
interacting fermions.

\vspace{0.1in}
\textbf{Acknowledgments}: V. G. would like to thank P. Bedaque for discussions. This research was supported by U.S.-ARO (S. T. \& V. G.),
ARO-JQI-MURI (B. M. A. \& C.-H. L.), ARO-DARPA-OLE (C.-H. L.) and NSF-JQI-PFC (C.-H. L.).

\appendix
\section{Explicit expression for the Green function matrix (\ref{GFmat})}
\label{GFexp}
Recall that the Green function matrix was given by $\cG(\bQ,\bk)=\check U\check D(\bQ,\bk)\check U^\dag$, where 
$\check U=\hat u_1\otimes\hat u_2$, and $\hat u_i$ and $\check D(\bQ,\bk)$ were defined in (\ref{ui}) and (\ref{diagmat}),
respectively. We then find
\begin{widetext}
\beq
\label{GFapp}
\cG(\bQ,\bk)=\frac{1}{4}\round{\begin{array}{cccc}
(++++) & ie^{-i\f_2}(+-+-) & ie^{-i\f_1}(++--) & -e^{-i(\f_1+\f_2)}(+--+)\\
-ie^{i\f_2}(+-+-) & (++++) & e^{i(\f_2-\f_1)}(+--+) & ie^{-i\f_1}(++--)\\
-ie^{i\f_1}(++--) & e^{i(\f_1-\f_2)}(+--+) & (++++) & ie^{-i\f_2}(+-+-)\\
-e^{i(\f_1+\f_2)}(+--+) & -ie^{i\f_1}(++--) & -ie^{i\f_2}(+-+-) & (++++)\end{array}},
\eeq
\end{widetext}
where $\f_i=\tan^{-1}(b_{y,i}/b_{x,i})$, and 
\beq
(p_1p_2p_3p_4)=\sum_{i=1}^4p_iD_i.
\eeq
Here, $p_i=\pm1$ and $D_i$ were defined in (\ref{D1})-(\ref{D4}).

\section{Self-consistency condition with a Zeeman field}
\label{zeemanapp}
In this appendix, we give a brief derivation of the self-consistency condition in the presence of a Zeeman term of the form
$H_Z=-\bH\cdot\hbbpm$, where $\hbbpm=(\hpmb,\hpm_z)$. The inverse single-particle Green function in this case can be written as
$\hat g^{-1}(\bk_i,\bQ)=\hI y(\bk_i,\bQ)+\al\hat{\bm\s}\cdot(\bbb_i\times\hat z)-H_z\hpm_z$, where
\beq
\bbb_i=\bb_i+\frac{1}{\al}\bH'=:(\bb'_i,b_{z,i}),
\eeq
where $\bH'=(-H_x,-H_y,H_z)$. It can be diagonalized with a unitary transformation
\beq
\hat u_i=\exp\curly{-\frac{i}{4}\square{\p+2\sin^{-1}\round{\frac{b_{z,i}}{\bar b_i}}}(\mathbf{n}'_i\cdot\hpmb)},
\eeq
where $\mathbf{n}'_i=\bb'_i/b'_i$.
By the composite unitary transformation $\check U=\hat u_1\otimes\hat u_2$ the two-particle inverse Green function
can be written as $\cG(\bQ,\bk)=\check U\check D(\bQ,\bk)\check U^\dag$, where $\check D(\bQ,\bk)$ 
is identical to (\ref{diagmat}) but with $b_i$ replaced by $\bar b_i$. The two-particle wave function coefficients
can be formed as before (c.f. (\ref{an})). The self-consistency condition then reads
\beq
\label{SC2D_genh}
\frac{1}{|V_0|}=
\int\frac{d^2\bk}{(2\p)^2}\frac{s\square{s^2-2\al^2\round{\bar b^2+\frac{\bar B^2}{4}}-f_1-f_2}}{d(\bQ,\bk)},
\eeq
where $\bar\bB=\bbb_1+\bbb_2$ and $\bbb=(\bbb_1-\bbb_2)/2$, and the denominator is defined as before 
(c.f. (\ref{denom})) but with $\bb$ and $\bB$ replaced by $\bbb$ and $\bar\bB$, respectively. $f_1$ and $f_2$
are given by
\begin{align}
f_1&=\frac{2\bar b_1\bar b_2}{b'_1b'_2}\square{\al^2\round{\bar b^2-\frac{\bar B^2}{4}}+H_z^2}\Theta_1\\
f_2&=-2\bar b_1\bar b_2\Theta_2,
\end{align}
where
\begin{align}
\Theta_1&=\frac{\sqrt{\abs{\bH'+\al^2\bbb}^2-H_z^2}\sqrt{\abs{\bH'-\al^2\bbb}^2-H_z^2}}
{\abs{\bH'+\al\bbb}\abs{\bH'-\al\bbb}}\\
\Theta_2&=\frac{H_z^2}{\abs{\bH'+\al\bbb}\abs{\bH'-\al\bbb}}.
\end{align}

\section{Obtaining $P(\bka_1,\bka_2)$}
\label{P12details}
$P(\bka_1,\bka_2)$ is given by the diagonal elements of the density matrix (\ref{dmgasa}). Since the momentum eigenstates
$\ket{\bq,\bka}$ are not eigenstates of the Hamiltonian, we must insert a resolution of unity using energy eigenstates. We
then obtain
\begin{multline}
\bra{\bq,\bka}e^{-\be\cH}\ket{\bq,\bka}=\\
\sum_n\int_{\bq'}\langle\bq,\bka|\psi_n(\bq')\rangle\langle\psi_n(\bq')|\bq,\bka\rangle e^{-e_n(\bq')/\tau}.
\end{multline}
At low temperatures, we can safely assume that most of the fermions are in the $n=0$ bound state. Therefore, we may approximate the
above matrix element by
\begin{multline}
\label{dmelement}
\bra{\bq,\bka}e^{-\be\cH}\ket{\bq,\bka}\approx\\
\int_{\bq'}\langle\bq,\bka|\psi_0(\bq')\rangle\langle\psi_0(\bq')|\bq,\bka\rangle e^{-e_0(\bq')/\tau}.
\end{multline}
Using (\ref{EE}) and (\ref{an}) we may write (\ref{dmelement}) as
\begin{multline}
\bra{\bq,\bka}e^{-\be\cH}\ket{\bq,\bka}\approx|c^0_\bq|^2V_0^2\langle\bq|\bq\rangle\\
\times\bra{00}\cG_0^\dag(\bq,\bka)\cG_0(\bq,\bka)\ket{00}e^{-e_0(\bq)/\tau}.
\end{multline}
(\ref{P12exp}) follows directly from this expression.


\end{document}